\documentclass[referee]{aa}
\usepackage{graphicx}
\usepackage{natbib} \bibpunct{(}{)}{;}{a}{}{,} 
\begin{document}

\title{Discovery of fourteen new ZZ Cetis with SOAR{\thanks{Based on observations at
the Southern Astrophysical Research telescope, a collaboration
between CNPq-Brazil, NOAO, UNC and MSU.}}
}

\author{
S. O. Kepler\inst{1,2},
B. G. Castanheira\inst{1},
M. F. O. Saraiva\inst{1,2},
A. Nitta\inst{3},
S. J. Kleinman\inst{3},
F. Mullally\inst{4},
D. E. Winget\inst{4},
D. J. Eisenstein\inst{5}
}

\offprints{kepler@if.ufrgs.br}
\institute{
Instituto de F\'{\i}sica, Universidade Federal do Rio Grande do Sul,
  91501-900  Porto-Alegre, RS, Brazil\\
\and SOAR, Casilla 603, La Serena, Chile\\
\and Sloan Digital Sky Survey, Apache Pt. Observatory, PO Box 59, Sunspot, NM 88349, USA\\
\and Department of Astronomy and McDonald Observatory,
  University of Texas,
  Austin, TX 78712, USA\\
\and Steward Observatory, University of Arizona\\
933 N. Cherry Ave.  Tucson, AZ 85721, USA\\
}

\date{Received --; accepted --}

\abstract{We report the discovery of fourteen new ZZ Cetis with the 4.1~m 
Southern Astrophysical Research telescope, at Cerro Pachon, in Chile.
The candidates
were selected from the SDSS (Sloan Digital Sky Survey) DA white dwarf stars
with
$T_{\mathrm{eff}}$ obtained from the
optical spectra fit, inside the ZZ Ceti instability strip.
Considering these stars
are multi-periodic pulsators and the pulsations propagate
to the nucleus of the star, they carry information on the
structure of the star and evolution of the progenitors.
The ZZ Cetis discovered
till 2003 are mainly within 100~pc from the Sun, and probe only
the solar vicinity. The recently discovered ones, and those
reported here, may sample a distinct population as they were
selected mainly perpendicular to the galactic disk and
cover a distance up to $\approx 400$~pc.
}

\titlerunning{SOAR ZZ Cetis}
\authorrunning{Kepler et al.}

\maketitle

\section{Introduction}
The ZZ~Ceti stars are pulsating white dwarf stars with an
atmosphere of pure hydrogen \citep{McGraw77, MR}. 
They show multi-periodic oscillations
with periods from 70s to 1500s and fractional amplitudes ranging
from 0.4\% to 30\%. They undergo $g$--mode pulsations caused
by the $\kappa$--$\gamma$ mechanisms and the development of
a sub-surface convection zone due to
the opacity bump caused by partial ionization of
hydrogen that starts when the cooling white dwarf reaches
effective temperatures around 12\,000K.
The convection zone stores and enhances
the heat exchange due to the pulsations.

The ZZ Ceti class of variable stars
is also called DAVs and is the coolest of the three
known instability strips in the white dwarf cooling sequence:
the pulsating
PG~1159 stars, around 200\,000--65\,000~K \citep{Dreizler98, 
Nagel,Quirion},
the DBV, around 25\,000--22\,000~K 
\citep{Beauchamp,Castanheira05a}
and the DAV,
around 12270--10850~K \citep{Bergeron04,Mukadam04}.

The ZZ Ceti class presents gradations between the two extremes:
the hot DAVs (hDAVs), close to the blue edge of the 
instability strip, have sinusoidal light-curves with low amplitude ($\leq$2\%)
and short periods ($\leq 300$\,s). The cool DAVs (cDAVs), close to the
red edge of the instability strip, show large amplitude ($\leq 30$\%)
long period pulsations ($\leq 1500$\,s), 
non-sinusoidal 
light curves
because they are distorted by the
extended convection zone \citep{Brickhill, Wu, Ising, Montgomery04}. Another important factor in shaping the
light curve
and defining which periods
are excited to observable amplitudes
is crystallization of the core \citep{W97, Montgomery99,
Metcalfe04, Kanaan05},
which for the high mass (above 1~$M_\odot$)
white dwarf stars occurs while the star is within the ZZ Ceti instability
strip, or before it reaches the strip, depending on the mass and the core composition. Pulsations cannot propagate inside a crystallized core, distorting
the period distribution and decreasing the pulsation amplitudes.

There are to date 89 known non-interacting ZZ Ceti stars
\citep{Fergal05,Castanheira05b}
among more
than 5400
spectroscopically identified
white dwarf stars \citep{MS}, but \citet{mcgraw77} and \citet{cox} already
indicated they are the most common variable star known. Because they
are intrinsically faint, $M_V \simeq 12$, the published
ones till 2003 are mainly within 100~pc from the Sun, and probe only
the solar vicinity. The recently discovered ones, and those
reported here, may sample a distinct population as they were
selected mainly perpendicular to the galactic disk and
cover a distance up to $\approx 400$~pc,
and the thin
disk scale height extends to $\approx 300$~pc \citep{MaS}.

White dwarf stars are the end points of evolution of stars
in the main sequence up to around 10.5~$M_\odot$
\citep[e.g.][]{Weidemann00}, i.e., close to
98\% of all stars. Taking into account the observed non-radial
$g$--mode pulsations \citep[e.g.][]{K84} are global pulsations,
with each pulsation mode constraining the stellar structure
in a different way, we can use the pulsations to untangle
the structure of the whole star \citep{W91, W94, K03,Metcalfe03} 
and even their rates of evolution
\citep{W85,Costa99,Costa03,K00,K05,Mukadam03}. 
These measured evolutionary rates have been used to calculate
the age of the coolest known white dwarf stars, allowing an
estimative of the age of the galactic disk \citep{W87}
and of a globular cluster \citep{Hansen02}.
Even more important,
pulsating white dwarf stars are excellent laboratories
for testing high energy and high density physics, such as
neutrino \citep{W04} and axion emission \citep{Corsico, K04}, 
crystallization \citep{W97, Montgomery03}, and even an estimation of 
${\mathrm{C(\alpha,\gamma)O}}$ reaction rate \citep{Metcalfe03a},
a rate important from early Universe composition to
supernova explosions, and which determines the size of the C/O core
of most white dwarf stars.
Crystallization, axion emission and cooling rates are mainly
determined from the study of the DAVs.
From evolutionary models, white dwarf stars with masses below 0.45~$M_\odot$
should have He cores, and those above 1.1~$M_\odot$ should have
O-Ne-Mg cores \citep[e.g.][]{Weidemann03}.
Another important use of pulsations is
to use the light travel time variations measurable by the phase changes in
the pulsation modes to detect planetary companions
to the white dwarf stars. As most planets will survive
post-main sequence mass loss to the white dwarf
phase \citep{Duncan,Mugrauer},
we can use the same technique used to study
companions in pulsars to detect even planets
smaller than the Doppler technique can,
complementing their search space. But planet searches
around white dwarf stars require very stable pulsations,
like those found in hot DAVs \citep{W03}, and
only a small sample of them is known to date.

We are therefore involved in a program to find a significant number
of pulsating white dwarf stars, to study their structure through
asteroseismology, measure their evolutionary rates, and look
for planets orbiting them.

\section{Candidate Selection}

The temperatures derived
from the optical spectra acquired by the Sloan
Digital Sky Survey and fitted to Koester's model
atmospheres \citep{Scot} are good selection criteria 
to choose candidates for time series photometric searches
of ZZ Ceti stars \citep{Mukadam04}.
The SDSS spectra have in general
SNR$\simeq 30$ for $g\simeq 18$ and we fitted Koester's
spectra models from 3800\,\AA\ to 7200\,\AA\ \citep{Scot}.
Unlike the fitting done by
\citet{Bergeron95,Bergeron04,KH}, which only fit the line profiles and not
the continuum, we used the whole spectra from 3800 to 7200\AA. 
The long wavelength baseline, coupled with the 
SDSS photometric data, and a low order multiplicative polynomial
to allow for small flux calibration uncertainties,
result in accurate $T_{\mathrm{eff}}$. The selection
of this limited wavelength range is to increase the
weight of the region with lines,  which are $\log g$
dependent.  
\citet{Mukadam04,Fergal05} and
\citet{Castanheira05b} show that we can attain 90\%
probability of variability if we constrain the
search range to 11800\,K $\geq T_{\mathrm{eff}} \geq$ 10850\,K.

\section{Observations}
We used the SOAR Optical Imager, a mosaic of two EEV
2048$\times$4096 CCDs, thinned and back illuminated,
with an efficiency around 73\% at 4000\,\AA,
to acquire time series photometry.  It covers
a field of 5.26'$\times$5.26' on the sky, on the bent cassegrain
port of the 4.1\,m SOAR telescope. We observed from March to May 2005,
when the telescope and imager were still under commissioning,
even lacking baffle tubes and therefore with an increased
background. We observed in fast readout mode, with
the CCDs binned 2$\times$2, which resulted in a pixel scale
of 0.154 arcsec/pixel and a readout+write
time of 10.2\,s. The exposure times ranged from 20\,s to 40\,s,
longer than the overhead  but still keeping the
Nyquist frequency in range with the shortest pulsation
periods detected to date. The data was bias subtracted and
flat fielded before we obtained differential photometry
through weighted apertures around 2\,FWHM 
(full width at half maximum of the seeing disk),
chosen for highest SNR. We observed each
star twice for around 2\,h each time.

All observations were obtained with a Johnson  B filter,
considering \citet{RKN} show the pulsation amplitude
increases to the blue, and to minimize the
background.

Table~\ref{t1} list the new variables and their 
effective temperatures obtained by fitting the
optical spectra to Detlev Koester's model
atmospheres, as in \citet{Scot}.
The first pulsator we observed was also observed with 
Argos \citep{Nather04} at McDonald Observatory 2.1~m telescope, to
confirm all the observed periodicities we detected, and check
validity of the whole observing system (the telescope, the instrument and
the software).

\begin{table}
\begin{minipage}{\linewidth}
\begin{center}
\label{t1}
\caption{New ZZ Cetis}
\begin{tabular}{lccccc}
\hline\hline
SDSS spSpec\\
MJD-Plate-Fiber&Name&g&$T_{\mathrm{eff}}$ (K)&log g&Main Periodicity
{\footnote{mma is milli-modulation amplitude, corresponding to
$1000\times \Delta F/F$, where $F$ is the measured flux.
The MJD-Plate-Fiber are the parameters necessary to
access the spectra at http://das.sdss.org.}}
\cr
\hline
52642-1185-085&WD 0825+0329& 17.48& 11801$\pm$ 105&8.33$\pm$ 0.044
          & 481s@4.5mma\cr
52650-1188-191&WD 0843+0431& 17.93& 11250$\pm$ 63&8.18$\pm$0.044
          & 373s@10.43mma\cr
52670-1190-322& WD 0851+0605& 17.08& 11306 $\pm$ 48&8.11 $\pm$ 0.029 
          &326s@22.4mma\cr
52238-0566-031& WD  0911+0310 & 18.41& 11634$\pm$ 126&8.11$\pm$ 0.084
          & 347s@17.4mma\cr
52976-1301-445& WD 0917+0926& 18.09& 11341$\pm$ 64&8.15$\pm$ 0.044
           & 289s@16.1mma\cr
51900-0278-367& WD 1106+0115& 18.37& 10990$\pm$ 62&8.09$\pm$ 0,049
           & 822s@12.2mma\cr
52672-1230-188& WD 1216+0922& 18.56& 11293$\pm$ 109&8.29 $\pm$ 0.078
          &823s@45.2mma\cr
52000-0288-412& WD 1218+0042& 18.71& 11123$\pm$ 93&8.16 $\pm$ 0.068
          & 258s@16mma\cr
52313-0333-077& WD  1222$-$0243& 16.74& 11398$\pm$ 44&8.35$\pm$ 0.026
          & 396s@22.0mma\cr
52026-0523-186& WD 1255+0211& 19.09& 11385$\pm$ 154&8.16 $\pm$ 0.106
          & 897s@31.7mma\cr
51689-0293-603& WD 1301+0107& 16.30& 11099$\pm$ 34&8.11 $\pm$ 0.023
           &879s@13mma\cr
51692-0339-629& WD 1310$-$0159& 17.67& 10992$\pm$ 65&7.92 $\pm$ 0.049
          & 280s@6.56mma\cr
51955-0298-608& WD 1337+0104& 18.57& 11533$\pm$ 156&8.55$\pm$ 0.085
           & 797s@10.2mma\cr
52045-0582-551& WD 1408+0445& 17.93& 10938$\pm$ 64&8.06$\pm$ 0.044
          & 849s@24.3mma\cr
\hline
\end{tabular}
\end{center}
\end{minipage}
\end{table}

\section{Results}
\begin{figure*} \centering \includegraphics[width=17cm]{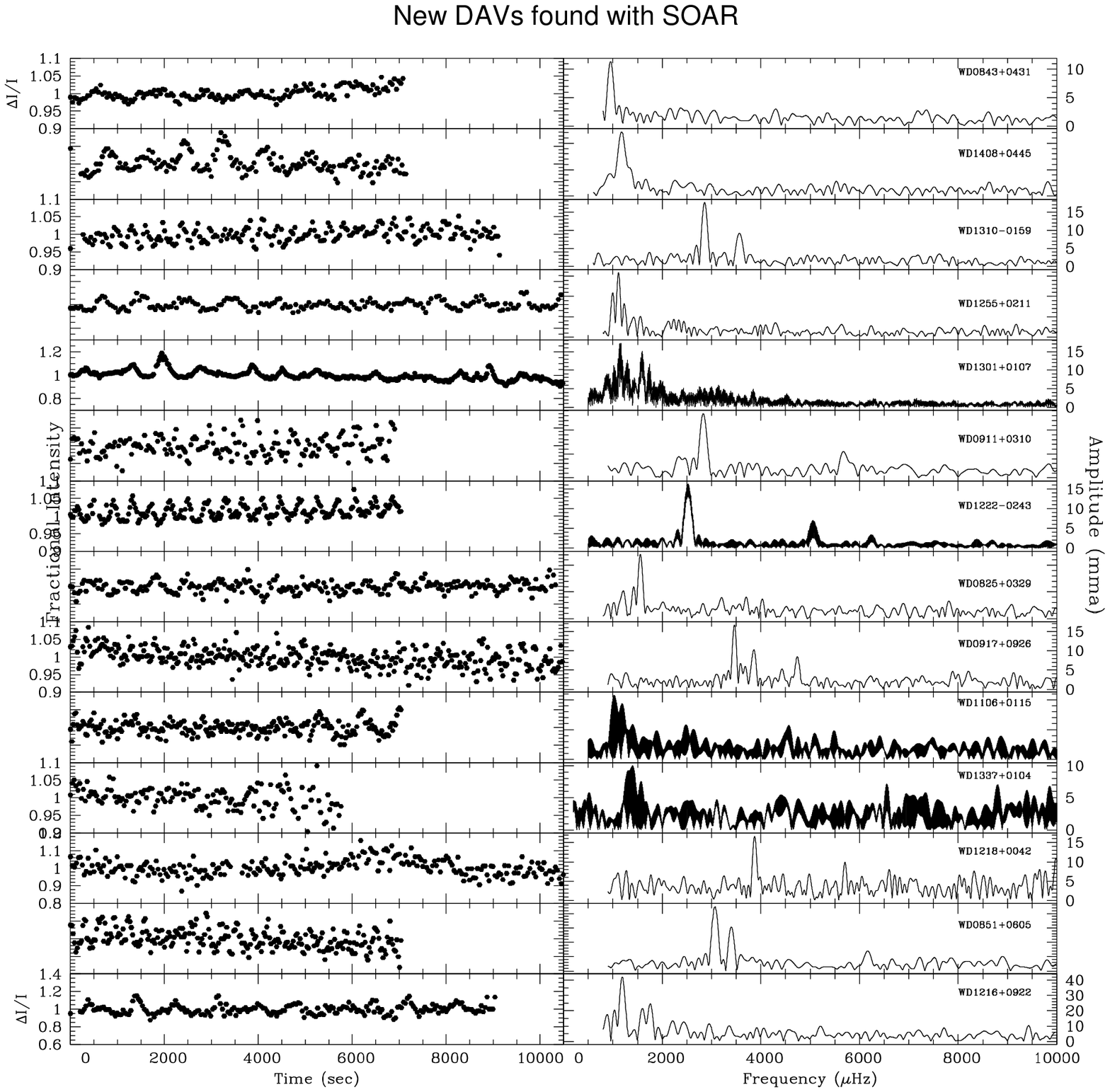} 
\caption{Light curves (left panels) and Fourier transforms (right panels) for 
the new ZZ Cetis. mma is milli-modulation amplitude, corresponding to
$1000\times \Delta F/F$, where $F$ is the measured flux.}
\label{lc}
\end{figure*}

In Fig.~\ref{lc}, we show the light curves on the left panels and the 
Fourier transform of them in the right panels.
In Table~\ref{t2} we list all runs obtained for each star and the
main periodicities detected,
i.e., those with a false alarm probability smaller than 1\%. 
$\langle A \rangle$ is the square root 
of the 
average power, and is an estimate of the noise
\citep{SC91,SC99,K93}.

\begin{table}
\begin{center}
\label{t2}
\caption{Periodicities detected in the light curves.}
\begin{tabular}{llcccc}
\hline\hline
Name&Mean noise&Period@Amplitude&Date of Obs.&Length\cr
\hline
WD 0825+0329&
          $\langle A\rangle$=1.6mma& 481s@4.5mma= 2.8$\langle A\rangle$& 10 Mar 05& 2h\cr
&&303s@3.8mma\cr
          &$\langle A\rangle$=3.37mma& 664s@10.79mma= 3.2$\langle A\rangle$& 11 Mar 05& 1h\cr
&&334s@6.9mma\cr
          &$\langle A\rangle$=2.13mma& 644s@12.0mma= 5.63$\langle A\rangle$& 13 Apr 05& 2.9h\cr
&&704s@6.0mma\cr
&&826s@5.3mma\cr

WD 0843+0431&
          $\langle A\rangle$=3.75mma& 373s@10.43mma=2.78$\langle A\rangle$&21 Mar 05& 2h\cr
	  &$\langle A\rangle$=2.02mma& 1049s@11.4mma=5.64$\langle A\rangle$&9 Apr 05& 2h\cr
          &$\langle A\rangle$=1.92mma& 1085s@7.42mma=3.86$\langle A\rangle$&11 Apr 05& 2.6h\cr

WD 0851+0605&
          $\langle A\rangle$=4.12mma&326s@22.4mma= 5.4$\langle A\rangle$&21 Mar 05& 2h\cr

WD 0911+0310 &
          $\langle A\rangle$=4.3mma& 347s@17.4mma&10 Mar 05&1.9h\cr
&&757s@16.4mma\cr
&&388s@12.3mma\cr
              &$\langle A\rangle$=5.1mma& 353s@26.9mma= 5.3$\langle A\rangle$&11 Mar 05& 1.9h\cr
&&176s@11.1mma\cr
              &McD $\langle A\rangle$=4.4mma& 352s@27.7mma= 6.2$\langle A\rangle$&11 Mar 05& 1.6h\cr
&&420s@12.6mma\cr

WD 0917+0926&
           $\langle A\rangle$=2.7mma& 289s@16.1mma= 6$\langle A\rangle$&14 Mar 05& 2.9h\cr
&&259s@10.2mma\cr
&&212s@8.0mma\cr
          & $\langle A\rangle$=3.0mma& 288s@14.0mma=4.6$\langle A\rangle$&15 Apr 05& 2h\cr

WD 1106+0115&
           $\langle A\rangle$=3.0mma& 822s@12.2mma= 4.1$\langle A\rangle$&15 Mar 05& 2h\cr
&&980s@10.1mma\cr
          & $\langle A\rangle$=3.9mma& 937s@11.1mma= 2.85$\langle A\rangle$&21 Mar 05& 2h\cr
&&719s@7.7mma\cr
&&220s@8.9mma\cr
&all data$\langle A\rangle$=2.5mma&973s@10.8mma\cr
&&842s@9.4mma\cr

WD 1216+0922&
          $\langle A\rangle$=10.7mma&823s@45.2mma= 4.2$\langle A\rangle$ &23 Mar 05& 1.9h\cr
&&409s@30.1mma\cr
          &$\langle A\rangle$=7.92mma& 840s@42.0mma=5.3$\langle A\rangle$& 13 Apr 05& 2.5h\cr
&&570s@24.6mma\cr
&&626s@21.6mma\cr
&&967s@20.5mma\cr

WD 1218+0042&
          $\langle A\rangle$=4.3mma& 258s@16mma=3.75$\langle A\rangle$&6 Apr 05& 3.4h\cr
&&175s@10.0mma\cr
&&100s@11.0mma\cr
          &$\langle A\rangle$=2.55mma&259s@8.2mma=3.2$\langle A\rangle$&15 Apr 05& 2h\cr
&&152s@5.1mma\cr
\hline
\end{tabular}
\end{center}
\end{table}

\begin{table}
\begin{center}
\label{t2a}
\caption{Periodicities detected in the light curves (cont.).}
\begin{tabular}{llcccc}
\hline\hline
Name&Mean noise&Period@Amplitude&Date of Obs.&Length\cr
\hline
WD 1222-0243& 
          $\langle A\rangle$=3.1mma& 396s@22.0mma= 7.1$\langle A\rangle$& 10 Mar 05& 2h\cr
&&198s@7.3mma\cr
          &$\langle A\rangle$=2.46mma& 198s@6.67mma= 2.7$\langle A\rangle$& 11 Mar 05& 1h\cr

WD 1255+0211&
          $\langle A\rangle$=4.84mma& 897s@31.7mma=6.55$\langle A\rangle$&9 Apr 05& 3.4h\cr
&&1002s@21.7mma\cr
&&812s@16.4mma\cr

WD 1301+0107&
           $\langle A\rangle$=3.8mma&879s@13mma= 3.4$\langle A\rangle$&11 Mar 05& 1.9h\cr
          & $\langle A\rangle$=7.8mma&    901s@24mma= 3$\langle A\rangle$&10 Mar 05& 0.6h\cr
          &     $\langle A\rangle$=4.4mma&    870s@22.3mma= 5$\langle A\rangle$&14 Mar 05& 2.9h\cr
&all data $\langle A\rangle$=3.0mma&882s@17.6mma\cr
&&628s@15.2mma\cr

WD 1310-0159&
          $\langle A\rangle$=2.92mma& 280s@6.56mma=2.25$\langle A\rangle$& 23 Mar 05&1.25h\cr
&&310s@6.4mma\cr
          &$\langle A\rangle$=2.81mma& 349.6s@17.6mma=6.3$\langle A\rangle$& 13 Apr& 2.5h\cr
&&280s@9.2mma\cr

WD 1337+0104&
           $\langle A\rangle$=3.7mma& 797s@10.2mma= 2.75$\langle A\rangle$&15 Mar 05& 2.5h\cr
          & $\langle A\rangle$=5.75mma&735s@18.49mma= 3.2$\langle A\rangle$&21 Mar 05& 1.6h\cr
&all data$\langle A\rangle$=3.0mma&715s@10.0mma\cr

WD 1408+0445&
          $\langle A\rangle$=4.34mma& 849s@24.3mma=$\langle A\rangle$& 15 Apr& 2h\cr
&&1038s@12.0mma\cr
&&764s@11.1mma\cr
\cr
WD 1359-0034&$\langle A\rangle$=0.69mma& NOV2&6 May& 2.2h\cr
&$\langle A\rangle$=2.4mma& &8 May& 1.1h\cr

WD 1432+0146\footnote{NOV5 Mukadam et al. (2004)}& 
          $\langle A\rangle$=1.25mma& NOV3&11 Apr& 3.6h\cr
\hline
\end{tabular}
\end{center}
\end{table}

We also observed the 
g=18.71 
variable discovered by \citet{Mukadam04},
WD~1502$-$0001, SDSS J150207.02$-$000147.1. 
Its SDSS spectrum spSpec-51616-0310-206 fits 
$T_{\mathrm{eff}}=11200\pm 117$, and $\log g=8.00\pm 0.079$.
with auto23.
Its
spSpec-51990-0310-229 spectrum fits
$T_{\mathrm{eff}}=11116 \pm 096$, and $\log g=8.18 \pm 0.06$,
with auto21. Auto23 is the version of the spectra fitting program
and calibration published by \citet{Scot}, while
\citet{Mukadam04} published values are older auto21.
\citet{Mukadam04} measured periodicities at:
687.5s@12.0\,mma,
629.5s@32.6\,mma,
581.9s@11.1\,mma,
418.2s@14.9\,mma, and
313.6s@13.1\,mma,
classifying it as a cDAV.

\begin{table}
\begin{center}
\label{t3}
\caption{Periodicities detected for the Variable WD 1502-0001.}
\begin{tabular}{llcccc}
\hline\hline
Name&Mean noise&Period@Amplitude&Date of Obs.&Length\cr
\hline
WD 1502-0001&
          $\langle A\rangle$=7.1mma& 603s@28.1mma=3.95$\langle A\rangle$&6 Apr 05& 2.7h\cr
&&658s@27.0mma\cr
&&415s@16.7mma\cr
&&141s@14.9mma\cr
          &$\langle A\rangle$=7.1mma& 604s@35.5mma=5$\langle A\rangle$&10 Apr 05& 2h\cr
&&424s@16.4mma\cr
\hline
\end{tabular}
\end{center}
\end{table}

We have also
found one star not observed to vary (NOV), WD1359-0034,
with a detection limit of $3\langle A \rangle=2$~mma. Its $T_{\mathrm{eff}}=10640 \pm 32$~K 
is outside the main strip found by \citet{Mukadam04}, but hotter than
their coolest variable.
We also confirmed one of the NOVs within the instability strip reported by 
Mukadam et al. 2004,
WD~1432+0146, with $T_{\mathrm{eff}}=11255\pm 73$\,K and $\log g=8.05\pm 0.05$,
at the detection limit of $3\langle A \rangle=4$~mma.

\section{Conclusions}
We fit the optical spectra acquired by SDSS with Koester's model
atmospheres, deriving the effective temperature of the DAs.
Selecting
to observe with time series photometry those inside the 
ZZ Ceti instability
strip derived by \citet{Mukadam04},
we detected fourteen new ZZ Cetis, i.e., hydrogen atmosphere
pulsating white dwarf stars, in the range
$11850~K \geq T_{\mathrm{eff}} \geq 10850~K$. 

We do note however that there
are 109 stars for which DR3 SDSS have multiple spectra,
just from 13000~K $\geq T_{\mathrm{eff}} \geq$ 10000~K, and the fitting
results
show that the 
mean uncertainties is 
$\sigma_{T_{\mathrm{eff}}}\simeq 300$~K, and 
$\sigma_{\log g}\simeq 0.21$~dex, for the same object but
different observations. This is
larger than
the internal uncertainty of the fits, 
but in general within $3\sigma$ of each other -- and 
mostly within 1 or 2$\sigma$, as in \citet{Scot}.
The uncertainties cover
a substantial fraction of
the instability strip. 
To really study the purity of the instability strip we need to
reduce the uncertainties to less than 200K, but we must take into
account the fact that the large amplitude
pulsators at the red edge have temperature excursions of around
500~K during one pulsation cycle \citep{RKN}. As the SDSS spectra
on average are $3 \times 900$s
exposures per observation,
it is unlikely that pulsations are
causing the 300K differences. 

\begin{figure}
\resizebox{\hsize}{!}{\includegraphics*[angle=0]{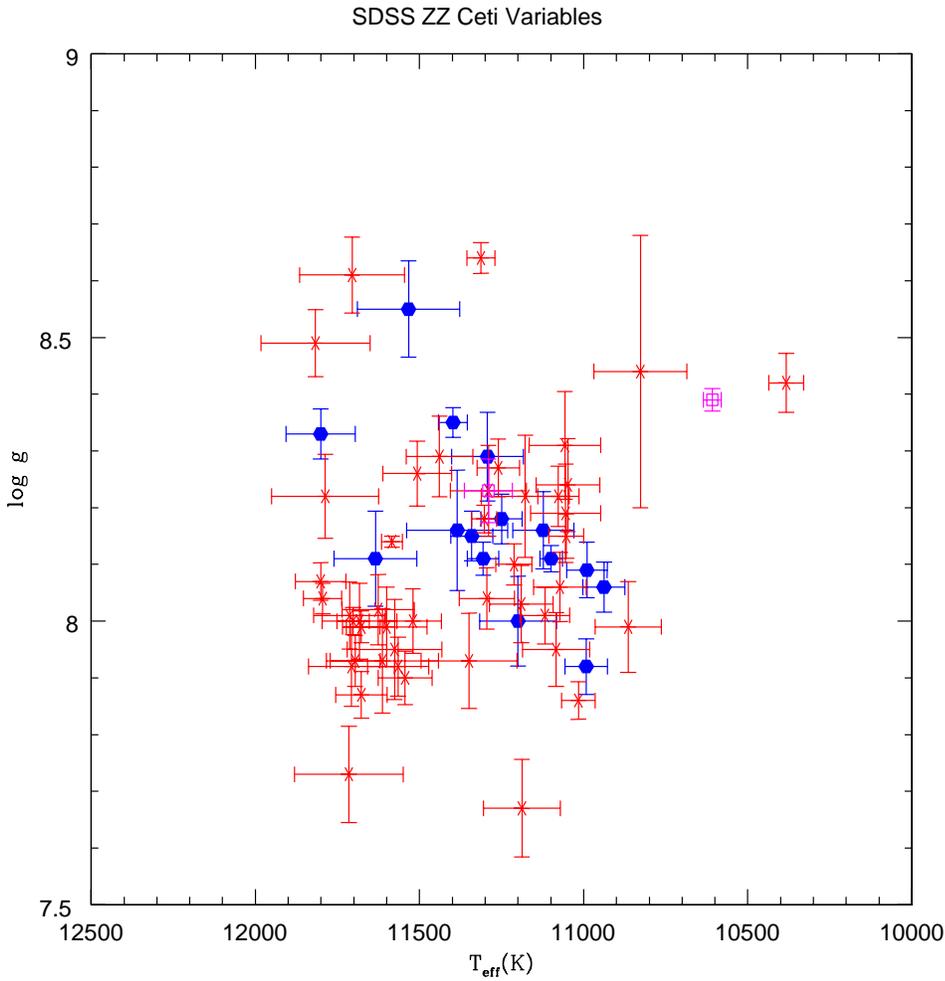}}
\caption{Plot of effective
temperature and log g of the observed objects in this paper
(filled symbols) and
SDSS DAVs in general (crosses). The two NOV studied
in this paper are represented by open rectangles.
}
\label{strip}
\end{figure}

\begin{acknowledgements}                                      
Financial support: NASA grant, CAPES/UT grant, CNPq fellowship.
\end{acknowledgements}                                        
                                                              
%
%
\bibliographystyle{aa} 

\end{document}